Marianne Gauffriau

Copenhagen University Library, The Danish Royal Library, Copenhagen, Denmark

Email: mgau@kb.dk

# Validation of counting methods in bibliometrics

## Introduction

The discussion about counting methods in bibliometrics is often reduced to the choice between full and fractional counting. Full counting gives the authors of a publication 1 credit each, while fractional counting shares 1 credit between the authors of a publication. However, several studies document that there are many counting methods and that the distinction between full and fractional counting is too simple (for examples, see Waltman, 2016, pp. 378–380; Xu et al., 2016).

The counting method may function as a core element in the definition of an indicator, for example, the *h*-index fractionally counted (Egghe, 2008). For some indicators, the counting method is the *only* element, for example, the number of publications (full counting) on a researcher's publication list.

The choice of counting method may affect the result of a bibliometric analysis that evaluates authors, universities, or other objects of study. The effect of shifting between counting methods can be seen in the *Leiden Ranking* where the user can choose between full and fractional counting for the indicators on *Scientific impact* (Centre for Science and Technology Studies - CWTS, 2019). A change from one counting method to the other alters the order of universities in the ranking.

Still, many studies do not justify the choice of counting method and there are no common standards for how to describe a counting method (Gauffriau, 2017, p. 678). The justification for the use of a specific counting method appear implicit in many bibliometric analyses. Thus, **the aim of the present study is to give an overview of counting methods in the bibliometric literature and to provide insight into their properties and use**.

Even though, the two studies mentioned in the beginning are quite comprehensive, the present study includes even more counting methods. I present a further development of the division into full, fractional, and other counting methods (Waltman, 2016, pp. 378–380) as I apply more and more well-defined categorizations of counting methods. A categorization based on data distributions (Xu et al., 2016) will not be investigated in the present study but may be applied in future research. In addition to the categorizations, I analyze the counting methods' internal and external validation.

The aim leads to **the research question (RQ): What types of counting methods are introduced into the bibliometric literature and how are they adopted in research evaluations?**

I operationalize the RQ through four sub-questions:

- RQ 1.1: How many counting methods are there in the bibliometric literature and when were each of the counting methods introduced into the bibliometric literature?

    RQ 1.1 is useful to understand the magnitude and timeliness of the RQ.

- RQ 1.2: To what extent can the counting methods introduced into the bibliometric literature be categorized according to general properties?

    Based on the overview provided through RQ 1.1, RQ 1.2 explores whether the counting methods share properties.

- RQ 1.3: To what extent are the counting methods introduced into the bibliometric literature validated internally?

    RQ 1.2 focuses on the development of categories and thereby general descriptions of counting methods. RQ 1.3 adds information about how the counting methods are validated internally by the studies that introduced the counting methods.

- RQ 1.4: How many of the counting methods introduced into the bibliometric literature are used in research evaluations and to what extent does the use in research evaluations - and thereby external validation - comply with the properties and internal validation of the counting methods?

    Whereas RQ 1.3 concentrates on the internal validation of the counting methods, RQ 1.4 investigates the external validation of the counting methods by how they are used in research evaluations. I investigate whether the intentions in the studies that introduced the counting methods (RQs 1.2 and 1.3) comply with the subsequent use of the counting methods.

## Methods

In RQ 1.1, I conduct a literature search in *Google Scholar* to identify counting methods and find out when they were introduced into the bibliometric literature. The literature search is restricted to peer reviewed studies in English from the period 1970–2018. A counting method used on different objects of study is only counted once. I do not include studies with minor variations of existing counting methods. Counting methods introduced after 2018 are not included as the use of these (RQ 1.4) would be difficult to assess less than two years after their introduction into the bibliometric literature.

In RQ 1.2, I apply two frameworks to categorize the counting methods (RQ 1.1) by their properties. The first framework describes selected mathematical properties of the counting methods (Gauffriau et al., 2007). The second framework is based on a qualitative text analysis and describes four groups of arguments for the choice of counting method

(Gauffriau, 2017).

In RQ 1.3, I assess the internal validation of the counting methods (RQ 1.1) following Gingras' three criteria for well-constructed bibliometric indicators: Adequacy, sensitivity, and homogeneity (Gingras, 2014, pp. 112–116). In the present study, the criteria are used on counting methods instead of indicators.

In RQ 1.4, I focus on the external validation by the use of the counting methods in research evaluations. In *Google Scholar*, I search for peer reviewed studies that present research evaluations and use the counting methods (RQ 1.1). I assess whether the use is consistent with the description of the counting methods (RQ 1.2) and with the internal validation of the counting methods (RQ 1.3).

Preliminary results

For RQs 1.1 and 1.2, I present selected results below. The research relating to RQs 1.3 and 1.4 is ongoing. Thus, I present examples from the preliminary findings.

RQ 1.1: The literature review identified 29 original counting methods. None of the counting methods was introduced into the bibliometric literature in the period 1970-1980. Seventeen were introduced in the period 2010-2018.

RQ 1.2: Twenty-seven of the 29 counting methods were categorized according to the selected mathematical properties. Thus, 21 are rank-dependent and fractionalized meaning that the authors of a publications share 1 credit but do not receive equal shares, for example harmonic counting (Hodge & Greenberg, 1981). Two counting methods are neither rank-dependent nor rank-independent.

For all 29 counting methods, the justification for their introduction into the bibliometric literature were categorized into the four groups of arguments for choosing a counting method. Twenty-five of the counting methods were introduced to support indicators that measure an object of study's contribution to, participation in, or production in a research endeavor.

RQ 1.3 (ongoing research): The first of Gingras' three criteria is whether an indicator is an adequate proxy for its purpose. Some of the methods used to determine the adequacy of a counting method imply circularity, which is not recommended by Gingras (Gingras, 2014, pp. 113–114). An example of a circularity is to compare scores obtained by the new counting method to scores obtained by an existing counting method (Stallings et al., 2013, p. 9681)[1]. Both counting methods build on publication counts. Instead, an independent and already accepted measure should be used (Gingras, 2014, pp. 113–114), for example, validate the scores of the new counting method for Nobel laureates versus other researchers (Aziz & Rozing, 2013, pp. 4–6)[1]. A difference in scores is expected due to the prestige of the laureates. Opposite to the counting method, prestige is not based on publication counts.

The second of Gingras' three criteria is the sensitivity of an indicator. The indicator should reflect the changes over time in the object the indicator is designed to measure (Gingras, 2014, p. 115). It is well-documented that co-authorship is the norm in most research fields and that the number of co-authors per publication is increasing over time (Henriksen, 2016;

---

[1] Note that this is a preliminary finding and serves as an example. Other studies apply the same method, and the cited study may apply other methods in addition to the one mentioned here.

Lindsey, 1980, p. 152; Wuchty et al., 2007). Thus, I assess how well the counting methods adapt to a change in the numbers of co-authors per publication. Some of the 29 counting methods have formulas that refer to surveys among researchers about co-author practice (Lukovits & Vinkler, 1995, p. 94)[1]. Eventually, these surveys may become obsolete due to changes in co-author practices and the counting method does not meet the sensitivity criteria.

The last of Gingras' criteria is the homogeneity of an indicator. Gingras advises to avoid a combination of publication and citation counts in the same indicator, for example the *h*-index. For such indicators it is not immediately visible whether a change in the score is due to a change in the publication count, citation count, or both (Gingras, 2014, p. 116). Some of the 29 counting methods do mix publication and citation counts (Shen & Barabasi, 2014, pp. 12325–12327)[1]. Many build on publication counts alone but add other elements such as parameters chosen by the bibliometrician (Kim & Diesner, 2014, pp. 591–593)[1] or separate formulas for groups of publications (Abramo et al., 2013, p. 201)[1]. In my analysis, these counting methods qualify as heterogeneous. In a research evaluation that uses a counting method with many elements, it may be difficult to see the effect of the different elements on the score obtained by the counting method.

RQ 1.4 (ongoing research): Only a few of the 29 counting methods are used in many research evaluations. Some counting methods are not used at all. For the counting methods that are used in research evaluations, I will analyze whether the use is consistent with the properties of the counting method (RQ 1.2) and with the internal validation of the counting method (RQ 1.3). In a previous study, I have shown that the use of common counting methods was not consistent across studies. One counting method was used for several purposes in the bibliometric literature (Gauffriau, 2017). The present study will show whether this observation holds for the 29 counting methods.

## Discussion and conclusion

The main RQ is: What types of counting methods are introduced into the bibliometric literature and how are they adopted in research evaluations? It is operationalized through four sub-questions (RQs 1.1-1.4).

The results under RQ 1.1 show that new counting methods keep being introduced into the bibliometric literature, hereof, 17 counting methods in the latest decade. These results supports the relevance and timeliness of the research question.

Furthermore, the results under RQ 1.2 demonstrate that it is possible create categories of counting methods that share the same properties. These categories pave the way for further analyses of the counting methods and for applying other well-defined frameworks for new categorizations.

The results relating to RQs 1.3 and 1.4 are not finalized. Still, preliminary results under RQ 1.3 based on Gingras' three criteria indicate that 1) a range of methods are used to determine the adequacy of the counting methods, 2) the evidence on which a counting method builds can make the counting method insensitive to changes in co-author practices, and 3) some counting methods are homogeneous while others comprise many elements and therefore are heterogeneous. These preliminary results indicate that the internal validation of counting methods is complex.

Under RQ 1.4, the external validation of the counting methods will be assessed by how the counting methods are used in research evaluations. So far, preliminary results document that not all counting methods are used in research evaluations.

The preliminary results of the present study show that counting methods in the bibliometric literature should not be reduced to a question about the choice between full and fractional counting. Previous studies have documented that many different counting methods exist in the bibliometric literature. Other studies have systematically analyzed a few counting methods using well-defined frameworks. In the present study, I do both. I identify 29 counting methods, which are categorized according to their properties and analyzed with regard to internal and external validation. This study has the potential to give a solid foundation for the use of and discussion about counting methods as well as encourages more analyses to develop our knowledge about counting methods.